\documentclass[conference]{IEEEtran}
\IEEEoverridecommandlockouts
\usepackage{cite}
\usepackage{amsmath,amssymb,amsfonts}
\usepackage{algorithmic}
\usepackage{graphicx}
\usepackage{textcomp}
\usepackage{xcolor}
\usepackage{psfrag}
\usepackage{float}

\begin{document}
\title{A Study on MIMO Channel Estimation by 2D and 3D Convolutional Neural Networks}


\author{\IEEEauthorblockN{Ben Marinberg}
\IEEEauthorblockA{\textit{Ben Gurion University} \\
\textit{Beer Sheva, IL}\\
benmar@post.bgu.ac.il}
\and
\IEEEauthorblockN{Ariel Cohen}
\IEEEauthorblockA{\textit{Ben Gurion University} \\
\textit{Beer Sheva, IL}\\
ariel5@post.bgu.ac.il}
\and
\IEEEauthorblockN{Eilam Ben-Dror}
\IEEEauthorblockA{\textit{Tel-Aviv Research Center} \\
\textit{Huawei Technologies Co. Ltd.}\\
eilam.ben.dror@huawei.com}
\and
\IEEEauthorblockN{Haim H. Permuter}
\IEEEauthorblockA{\textit{Ben Gurion University} \\
\textit{Beer Sheva, IL}\\
haimp@bgu.ac.il}
}
\maketitle
\begin{abstract}
In this paper we study the usage of Convolutional Neural Network (CNN) estimators for the task of Multiple-Input-Multiple-Output Orthogonal Frequency Division Multiplexing (MIMO-OFDM) Channel Estimation (CE). Specifically, 
the CNN estimators interpolate the channel values of reference signals for estimating the channel of the full OFDM resource element (RE) matrix.
We have designed a 2D CNN architecture based on U-net, and a 3D CNN architecture for handling spatial correlation. We investigate the performance of various CNN architectures for a diverse data set generated according to 5G NR standard, and in particular we investigate the influence of spatial correlation, Doppler and reference signal resource allocation. The CE CNN estimators are then integrated with MIMO detection algorithms for testing their influence on the system level Bit Error Rate (BER) performance.
\end{abstract}

\vspace{-0.1cm}
\section{Introduction}

MIMO-OFDM is a fundamental technology in 4G and 5G standards for wireless communications, which achieves high spectral efficiency and enables the ever-growing demand for data throughput and capacity. Efficient data detection at the receiver is highly dependent on accurate representation of the fading channel coefficients. Channel estimation is the process of calculating these coefficients, and it usually involves the transmission and detection of predefined reference signals (RS, a.k.a pilots). Each RS can be used for calculating the channel response at its allocated Resource Element (RE) by applying the Least Squares 
algorithm. However, the overhead of allocating REs for reference signals reduces the number of REs available for data transmission. As the MIMO order increases towards massive MIMO, the problem of performing accurate channel estimation with a minimal amount of allocated RS becomes harder. 

Traditionally, wireless communication networks have been designed according to explicit mathematical models, and researchers have developed model-driven algorithms which aim to represent and handle practical field conditions. In recent years, as machine-learning (ML) techniques improved, much research has been dedicated to data-driven algorithms, which do not assume a predefined model, but rather learn from provided data samples. The authors in \cite{modelorai} elaborate on the integration of model-based and AI-based approaches for wireless networks, and a comprehensive survey of recent advances and future challenges for ML application to wireless networks is provided in \cite{ai4b5g}. 

\subsection{Model Driven Channel Estimation Algorithms}
Consider an OFDM subframe with $N_{sc}$ frequency subcarriers and $N_{symb}$ time symbols, and a MIMO setup of $N_T$ transmitters and $N_R$ receiver antennas. For each RE the received signal can be modeled in the frequency domain as:
 \begin{equation}\label{MIMO_eq}
     Y_{f,s}=H_{f,s}\cdot X_{f,s} + W_{f,s},
 \end{equation}
where $f\in\{1,..,N_{sc}\}$ is the frequency subcarrier index and $s\in\{1,..,N_{symb}\}$ is the time symbol, $Y_{f,s}\in \mathbb{C}^{N_R\times 1}$ is the received signal, $H_{f,s}\in \mathbb{C}^{N_R\times N_T}$ is the channel response matrix, $X_{f,s}\in \mathbb{C}^{N_T\times1}$ is the transmitted signal, and $W_{f,s}\in \mathbb{C}^{N_R\times1}$ is additive white complex Gaussian noise. The purpose of channel estimation is to calculate $H_{f,s}$ for correctly detecting the transmitted signal $X_{f,s}$ given the received signal $Y_{f,s}$.  

Least Squares (\textbf{LS}): Given a RS at subcarrier $f_p$ and symbol $s_p$, the LS estimation of the channel matrix $\hat{H}^{LS}_{f_p,s_p}$ is defined as:
\begin{equation}
    \hat{H}_{f_p,s_p}^{LS}=Y_{f_p,s_p}\cdot\ (X_{f_p,s_p}^{-1})^T.
\end{equation}
Next, an estimation is required for REs with no RS. The following model-driven algorithms, based on \cite{cho}, interpolate the LS results, and will be used for comparison:
\begin{enumerate}
\item Linear Interpolation (\textbf{LI}): averaging the LS results over all time symbols, then applying a linear interpolation between each pair of pilot REs in the frequency axis, and replicating the result vector for every time symbol.
\item DFT-based Interpolation (\textbf{DFTI}): averaging the LS results over all time symbols, then decreasing the noise by applying DFT and eliminating all impulse response beyond the maximum channel delay. Then replicating the result vector for every time symbol.
\item DFT-Linear Interpolation (\textbf{DFTLI}): perform DFTI for each OFDM symbol in which pilots are transmitted, and then perform linear interpolation in the \emph{time} axis for OFDM symbols with no pilots. This algorithm is designed to follow fast-fading channels.
\end{enumerate} 
Additional model-driven CE algorithms such as Minimum Mean Square Error (MMSE) are described in \cite{cho,mmse_est}. 
The MMSE estimator relies on having the second order channel statistics, i.e., the auto-correlation matrix. In this paper we assume that the channel estimation is done from a given resource block of pilot symbols without any prior knowledge, hence MMSE is not suitable. In future work, as described in Section \ref{s_conclusion_future} we will analyse also estimation algorithms based on long-time sequential of data and there MMSE will be considered.  

\subsection{Data Driven Channel Estimation Algorithms}
In \cite{MMSE-CE} the authors present a CNN-based low complexity estimator, which is motivated by the structure of MMSE for channels which satisfy the Toeplitz assumption. This CNN estimator is used in \cite{antennaarrays} for investigating its performance with various antenna array configurations. In \cite{doublyselective} a Deep Neural Network (DNN) is suggested for frequency and time selective (doubly selective) fading channels, using the results of the LS algorithm and the estimated channel of the previous block as inputs to the DNN. In \cite{untrained} a DNN that does not require training is used for de-noising the received signal, followed by conventional LS estimation. The authors in \cite{realistic} propose a neural network (NN) for realistic channel modeling, which can be used for mitigating pilot contamination and for channel compressing and fingerprinting. In \cite{mumble} they use sequence-to-sequence learning models for performing channel prediction. Another channel prediction example is demonstrated in \cite{aging}, where 
CNN Auto-Regressive (CNN-AR) and CNN Recurrent Neural Network (CNN-RNN) 
architectures are used for Channel State Information (CSI) forecasting by taking into account channel aging features. An end-to-end approach is taken in \cite{pilotdesign} for jointly designing the pilot signals and the channel estimator using DNNs.

\subsection{Main Contributions}
In this paper we explore the channel estimation performance of several CNN-based architectures under various channel fading models, Doppler shift values, antenna correlation levels and signal-to-noise ratios (SNR). We propose a unique 2D U-net estimator as well as a 3D feed-forward estimator, which exploits the extra dimension for handling spatial correlation. The proposed CNN estimators outperform the above reference model-based estimators without any prior knowledge of the aforementioned channel parameters. We further investigate the estimators' performance for different RS allocation schemes. Finally, we integrate the CE CNN estimators with MIMO detection algorithms and present the system-level performance in terms of bit-error-rate.

\section{Problem Definition}
As discussed above, the purpose of channel estimation is to calculate the channel response for every RE and for every pair of Tx-Rx antennas, so $H\in \mathbb{C}^{N_{sc}\times N_{symb}\times N_R\times N_T}$ is 4-dimensional. LS is used for calculating the channel response for pilot REs. Then, a CE algorithm should further apply a mechanism for filling in the sparse 4D $H$ matrix. 
\begin{figure}[h!]
\begin{center}
\begin{psfrags}
    \psfragscanon
    \psfrag{A}[]{Dense allocation} 
    \psfrag{B}[]{Sparse allocation}
    \psfrag{C}[c][c][1][90]{Frequency}
    \psfrag{D}[b]{OFDM symbol}
    \psfrag{a}[]{ \tiny{Tx1}}
    \psfrag{b}[]{ \tiny{Tx2}}
    \psfrag{c}[c]{ \tiny{Tx3}}
    \psfrag{d}[c]{ \tiny{Tx4}}
    \psfrag{e}[c]{ \tiny{Tx5}}
    \psfrag{f}[c]{ \tiny{Tx6}}
    \psfrag{g}[c]{$\ $\tiny{Tx7}}
    \psfrag{h}[c]{ \tiny{Tx8}}
\includegraphics[width=8cm]{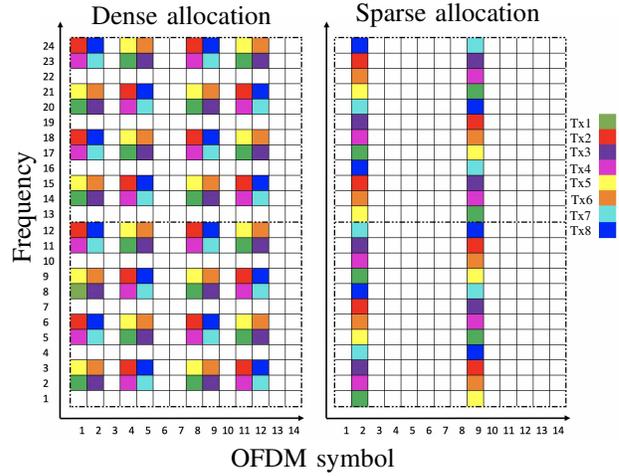}
\vspace{-0.3cm}

\caption{RS allocation types of 8 transmitting antennas. \vspace{-0.9cm}} \label{RS}
\psfragscanoff
\end{psfrags}
\end{center}
\vspace{-0.7cm}
\end{figure}

Reference signals can be allocated over the OFDM time-frequency grid in many ways. Fig. \ref{RS} depicts two possible allocations for a 24-subcarriers over 14-symbols 8x8 MIMO subframe: sparse and dense. The dense RS allocation has 16 RS for each transmitting antenna, resulting in a signaling overhead of 34\% of the REs. Sparse allocation has 6 RS for the same configuration, which reduces the signaling overhead to 14\% of the REs. 




We measure CE performance according to two criteria: Normalized Mean Square Error (NMSE) 
of the channel estimator output in decibels, and the resulting Bit Error Rate (BER) of the MIMO detection algorithm. For a channel grid $H$ and its estimation $\hat{H}$ the NMSE 
criterion in decibels is defined as: 
\begin{equation}
  \text{NMSE}\left( H,\hat{H}\right) =10\cdot \log_{10}\left( \frac{ ||H-\hat{H}||_F^2 } { ||H||_F^2 } \right),  
\end{equation}
where $||\cdot||_F^2$ is the square of the Frobenius norm. Frobenius norm is defined as the square root of the sum over squared absolute value of all input elements.

We test our suggested solution under various channel conditions, including several spatial correlation and Doppler shift values. 
Spatial correlation is defined as the correlation measured between each pair of antennas. For a correlated channel matrix $H_{f,s}$ and correlation matrices 
$R_R,R_T$ of Rx and Tx respectively, the spatial correlation can be formulated as in \cite{corr}:
\begin{equation}
    H_{f,s}=R_R^{\frac{1}{2}}\cdot H_{f,s}^{iid} \cdot R_T^{\frac{1}{2}},
\end{equation}
where $H_{f,s}^{iid}$ is a matrix of a non correlated channel and its elements are independent and identically distributed. Doppler shift is a frequency shift due to the mobility of either the transmitter, the receiver or both. 

\begin{figure}[h]
\begin{center}
\begin{psfrags}
    \psfragscanon

        \psfrag{a}[l][][0.55]{Layer $1$}
        \psfrag{b}[l][][0.55]{Layer $2$}
        \psfrag{c}[l][][0.55]{Layer $7$}
        \psfrag{d}[l]{$\hdots$ }
        \psfrag{G}[][][0.4]{$\quad\ \ \ $GELU}
        \psfrag{Y}[][][0.6][90]{1320}
        \psfrag{Y}[][][0.6][90]{1320}
        \psfrag{R}[b][][0.8]{$\operatorname{Re}$}
        \psfrag{I}[b][][0.8]{$\operatorname{Im}$}
        \psfrag{T}[t][][0.6]{14}
        \psfrag{P}[b][][0.8]{\ \ \ \ \ \ \ \ \ Sparse LS Input}
        \psfrag{L}[b][][0.8]{\ \ \ \ \ \ Full Output}
\includegraphics[width=9cm]{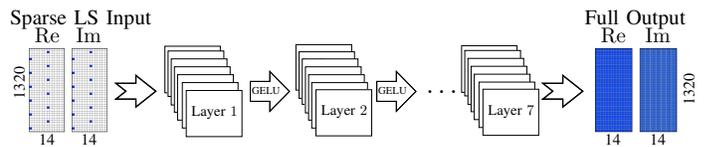}
\caption{2D FF CNN scheme for CE \vspace{-0.8cm}} \label{2DCNN}
\psfragscanoff
\end{psfrags}
\end{center}
\end{figure}

\section{Proposed CNN Estimators}
CNN models are widely used for image processing and image in-painting tasks (cf. \cite{inpaint}) in which pattern recognition is required. The sparse 4D channel matrix which contains the RS channel values bears resemblance to images, as both are characterized by correlation of adjacent elements. Therefore, our proposed CNN estimators were inspired by image processing algorithms such as super resolution and in-painting networks.

A CNN consists of convolution layers and activation functions. The input of a 2D-CNN architecture is a two dimensional matrix, as depicted in Fig. \ref{2DCNN}, whereas the input of a 3D-CNN is a three-dimensional matrix. 
In a simple feed-forward (FF) CNN architecture the layers are serially connected, so that each layer's output serves as the input of its successor. 

\begin{figure}[h!]
\begin{center}
\begin{psfrags}
    \psfragscanon
        \psfrag{A}[][][0.7]{Layer $1$}
        \psfrag{B}[][][0.7]{Layer $2$}
        \psfrag{C}[][][0.7]{Layer $3$}
        \psfrag{D}[][][0.7]{Layer $4$}
        \psfrag{E}[][][0.7]{Layer $5$}
        \psfrag{F}[][][0.7]{Layer $6$}
        \psfrag{G}[][][0.7]{Layer $7$}
        \psfrag{f}[b]{Feed-Forward}
        \psfrag{u}[b]{U-net}
        \psfrag{d}[b]{DenseNet}
        \psfrag{r}[b]{ResNet}
\includegraphics[width=8cm]{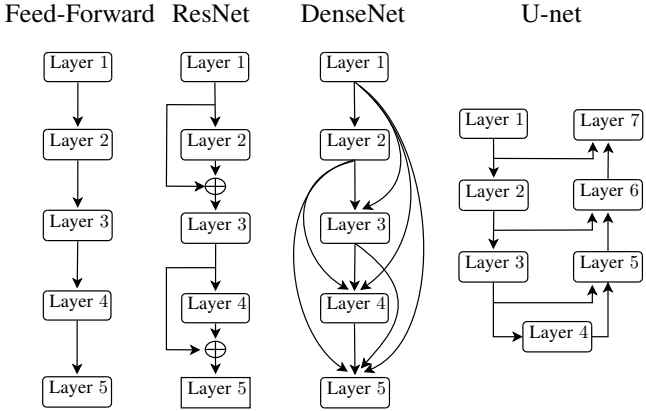}
\caption{Inter-layer connectivity schemes investigated in this paper} \label{interlayers}
\psfragscanoff
\end{psfrags}
\end{center}
\end{figure}

Other CNN architectures apply more complex inter-layer connectivity for improving the gradient flow, increasing the model's generalization capacity, and bypassing saturated weights and layers. As depicted in Fig. \ref{interlayers}, a ResNet \cite{resnet} architecture adds the current layer's output to the output of the previous layer for every other layer. In the U-net\cite{unet} architecture layers are symmetrically inter-connected, so that the output of the first layer feeds both the second layer and the last layer, and so on. A DenseNet \cite{dense} architecture takes this a step further, and connects the output of each layer to all of its succeeding layers.

\vspace{-0.4cm}
Having tested various configurations of 2D and 3D CNNs with FF, U-net and DenseNet connectivity, we found that the 2D U-net (\textbf{2DU}) and the 3D feed-forward (\textbf{3DFF}) architectures had superior CE performance over the other configurations. We will thus discuss the details of these two architectures.
\vspace{-0.2cm}
\subsection{Common Features}
Both estimators have non-linear activation functions for all layers but the last. We tested sigmoid, ReLU and GELU \cite{gelu} activation functions, and found that GELU activation provides the best results. A GELU activation function is defined as 
\vspace{-0.1cm}
\begin{equation}
    \text{GELU}(x)=x\cdot \Phi(x),
    \vspace{-0.1cm}
\end{equation}
where $x$ is the input and $\Phi(x)$ is a standard normal distribution CDF.
We set the loss function to be Minimum Square Error (MSE) loss and added $\text{L}_2$ regularization to prevent over-fitting. Denoting $x$ as an input sample, $\hat{H}(x)$ as the estimator's output, $H$ as the target output and $\Theta$ as the model's weights, we define the loss function as:
\vspace{-0.15cm}
\begin{equation}
    Loss\left(H,\hat{H}(x)\right) = ||H-\hat{H}(x)||_F^2 + \lambda ||\Theta||_F^2,
    \vspace{-0.15cm} 
\end{equation}
where $\lambda$ is a hyper-parameter for tuning the balance between $\text{L}_2$ regularization and the MSE. We use AdamW optimizer \cite{adamw}, which is designed to improve gradients when using $\text{L}_2$. Batch normalization and dropout did not improve the performance, so we have not used them in our final estimators.

\subsection{2DU Estimator}
The 2DU estimator 
uses the LS estimation of the RS in 
the frequency and time dimensions 
for interpolating
the channel of each Tx-Rx antenna pair separately. This means that ${N_T\times N_R}$ instances of the 2DU estimator should be used for every CE subframe.
A single model is trained for all Tx-Rx pairs.
Using U-net inter-layer connectivity has improved the 2D FF CE performance without changing the layers' dimensions. The structure of the 2DU estimator is described in Table \ref{structure_table}.

\subsection{3DFF Estimator}
The 3DFF estimator uses 3D convolution kernels which are applied to the input's frequency, time and Rx antennas dimensions. This property allows the 3DFF estimator to handle 
Rx
spatial correlation. Its input is the LS results of all $N_{R}$ Rx antennas for a given Tx antenna, so $N_T$ instances of 
a single-trained 
3DFF estimator model should be used for each CE subframe. The estimator's structure is described in Table \ref{structure_table}.
\vspace{-0.4cm}
\begin{table}[htbp] 
    \begin{center}
    \caption{2DU and 3DFF Estimator Structure}
\begin{tabular}{c||c|c|c|c||c}
    &  \multicolumn{2}{c|}{2DU} & \multicolumn{2}{c||}{3DFF} & \\\cline{2-5}
     Layer&Kernel&In chan.&Kernel&In chan.&Out chan.\\\hline\hline
     1&$7\times5$&2 & $7\times5\times5$ &  2& 10  \\\hline
     2&$7\times5$&10 & $7\times5\times5$ & 10 & 10 \\\hline
     3&$7\times5$&10 & $7\times5\times5$ & 10 & 10 \\\hline
     4&$7\times5$&10 & $7\times5\times5$ & 10 & 10 \\\hline
     5&$5\times5$&20 & $5\times5\times5$ & 10 & 20 \\\hline
     6&$5\times5$&30 & $5\times5\times5$ & 20 & 5  \\\hline
     7&$3\times3$&15 & $3\times3\times3$ & 5 & 2\\
\end{tabular}
\label{structure_table}
\end{center}
\end{table}
\vspace{-0.4cm}

An important consideration when designing a neural network for wireless communications is its complexity, which should allow the network's inference to meet strict real-time requirements. Comparing the complexity of the proposed estimators for 8x8 MIMO, 2DU has 25.2K parameters, whereas 3DFF has 93.7K parameters. The total amount of floating point operations (FLOPS) per subframe is 467M and 2304M FLOPS for 2DU and 3DFF respectively, so 3DFF requires 5 times more FLOPS than 2DU. 
This complexity is much higher than that of LI, DFTI and DFTLI, which require 4K, 1M and 3.5M FLOPS, respectively. 
However, using the MMSE estimation in \cite{mmse_est} with LI requires 14.3G FLOPS, which is 6 times more complex than 3DFF.
By using inference accelerators or processors which support Single Instruction Multiple Data (SIMD) parallelization, real-time computation of 2DU and 3DFF may be maintained.


\section{Data Generation}
We have used MATLAB\textsuperscript{\textregistered} 5G Toolbox\textsuperscript{\texttrademark} which is 5G New Radio standard compliant for creating our data sets.
We used an OFDM 1ms subframe grid of 1320 subcarriers over 14 symbols, and applied either a dense or a sparse RS allocation pattern onto it. In addition, we used a broad range of channel conditions for having a diverse data set, aiming to achieve a single generalized estimator that copes with any combination of these parameters. 

Specifically, our test set includes 3 spatial correlation levels of the Rx antennas: low, medium and high, as defined in \cite{corr} and 5 fading models as defined in \cite{TDL}. We used Doppler shift values in the range of 0Hz to 120Hz and SNR levels in the range of -10dB to 30dB. 
A single training set consists of 160k samples and a test set has 80K samples, where the shape of each sample is [1320-subcarriers, 14-symbols, 2-complex-parts].

\section{Results and Insights}
All results presented in Figures \ref{8x8_8p_2d} to \ref{8x8_dops} relate to a single training-set for 8x8 MIMO with dense RS allocation, so we use one trained model for all fading models, spatial correlations, Doppler shifts and SNR values.
 \subsection{2D and 3D CNN Estimators}
 \begin{figure}[h]
\begin{center}
\vspace{-0.6cm}
\begin{psfrags}
\includegraphics[trim={0.1cm  0.3cm  0.1cm 1.2cm},clip,width=9cm]{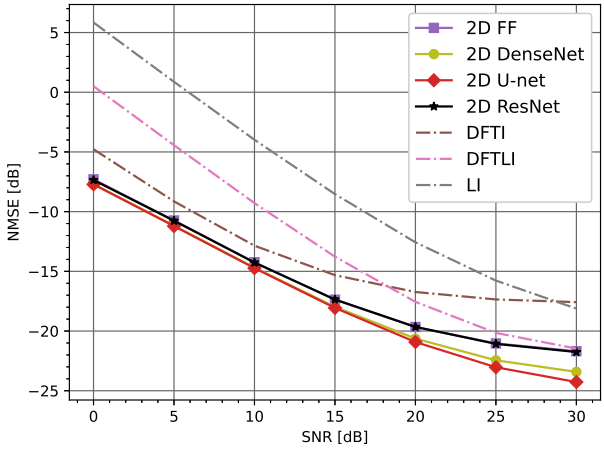}
\caption{\textbf{2D CNN}: CE performance for 8x8 MIMO, dense RS. 2DU performs better than the other 2D CNN (2D FF, 2D DenseNet, 2D ResNet) and better than model-driven estimators (DFTI, DFTLI, LI).\vspace{-0.3cm}} \label{8x8_8p_2d}
\psfragscanoff
\end{psfrags}
\end{center}
\end{figure}

In Fig. \ref{8x8_8p_2d} we compare 2D CNN performance for all 4 inter-layer connectivity schemes, presenting the NMSE 
averaged over all spatial correlation, fading model and Doppler shift values.
It can be clearly seen that all 2D CNN (2D FF, 2D DenseNet, 2D U-net, 2D ResNet) estimators perform better than all model-driven estimators (DFTI, DFTLI, LI).
The 2D FF estimator benefits from increasing the inter-layer connectivity, and the optimal complexity balance is achieved by the 2DU architecture. Increasing the connectivity to a Dense-Net does not improve the results. 
\begin{figure}[h!]
\begin{center}
\begin{psfrags}
    \psfragscanon
\includegraphics[trim={0.1cm  0.3cm  0.1cm 1.2cm},clip,width=9cm]{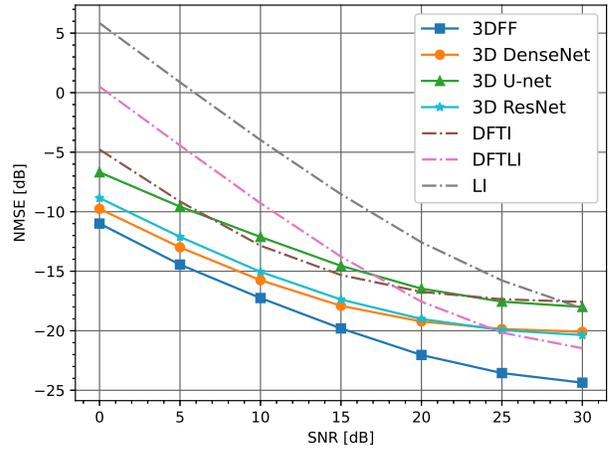}
\caption{\textbf{3D CNN}: CE performance for 8x8 MIMO, dense RS. 3DFF is clearly better than all other estimators.\vspace{-0.3cm}} \label{8x8_8p_3d}
\psfragscanoff
\end{psfrags}
\end{center}
\end{figure}

Next, in Fig. \ref{8x8_8p_3d} we compare the performance of 3D CNN architectures for the same data set. Here, the 3DFF estimator performs better than all other estimators, as it is complex enough to achieve good generalization without further inter-layer connectivity, which actually deteriorates the performance
due to more complex training convergence.

\subsection{Spatial Correlation and Doppler Effect}
\begin{figure}[h!]
\begin{center}
\vspace{-0.3cm}
\begin{psfrags}
    \psfragscanon
\includegraphics[trim={0.1cm  0.3cm  0.1cm 0.9cm},clip,width=9cm]{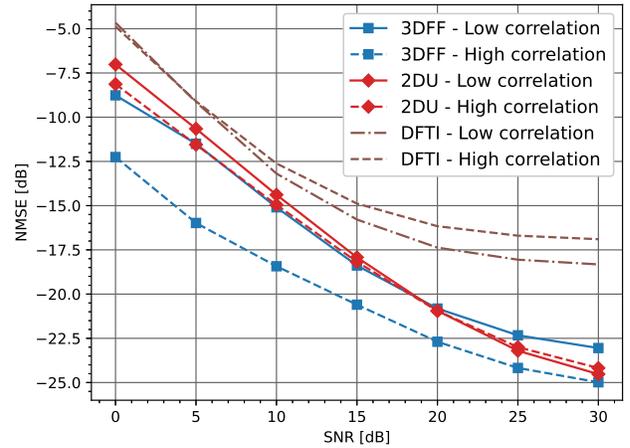}
\caption{\textbf{Rx Spatial Correlation}: CE performance for low and high correlation. The 3DFF estimator performs much better for high correlation, whereas 2DU is preferable for low correlation due to its lower complexity. Both CNN estimators perform better than model-driven estimators regardless of the correlation level.\vspace{-0.1cm}} \label{8x8_8p_cors}
\psfragscanoff
\end{psfrags}
\end{center}
\end{figure}

\begin{figure}[h!]
\begin{center}
\begin{psfrags}
    \psfragscanon
\includegraphics[trim={0.1cm 0.3cm  0.1cm 1.2cm},clip,width=9cm]{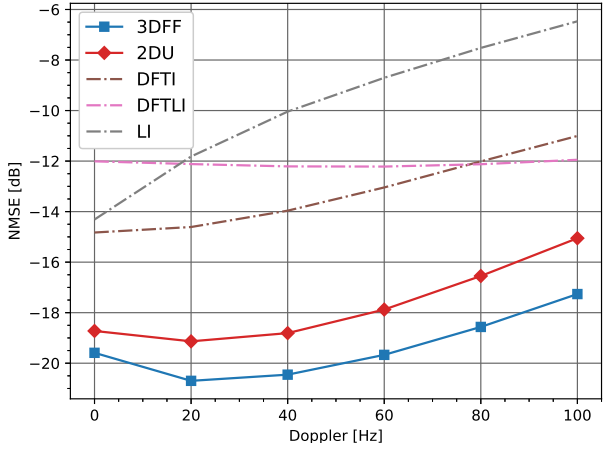}
\caption{\textbf{Doppler shift}: CE performance as a function of Doppler shift. Note that the performance of both CNN estimators for high Doppler shift is even better than the model-driven estimators at low Doppler shift.\vspace{-0.4cm}} \label{8x8_dops}
\psfragscanoff
\end{psfrags}
\end{center}
\end{figure}

We select the 2DU and the 3DFF estimators which achieved the best results, and in Fig. \ref{8x8_8p_cors} we compare their performance under low and high spatial correlation of the Rx antennas.
\vspace{-0.2cm}
As expected, both estimators have similar results for low spatial correlation, while the 3D estimator is much better for high correlation, as it exploits its extra dimension of Rx antennas. 
For high SNR, however, this advantage is redundant, and the lower complexity of the 2DU estimator is favorable.
In Fig. \ref{8x8_dops} we examine the effect of Doppler shift on the 2DU and 3DFF estimators, averaged over all spatial correlations, fading models and SNR values. Both CNN estimators significantly outperform the model-driven estimators. 
Furthermore, since we train our models using all Doppler values, they perform well with all of them.
On the other-hand,  LI and DFTI, which assume a slow fading model, perform better than DFTLI  at low Doppler shift values. As expected, the performance of the DFTLI estimator is similar for all Doppler values, as it assumes a fast fading channel.

\subsection{RS Resource Overhead Reduction}
A key advantage of using a data-driven estimator is the possible reduction of RS resource allocation overhead. 
We have trained the 3DFF estimator with a sparse RS pattern data set, and compared the results to those of the dense RS allocation data set for 8x8 MIMO. Recall that the dense allocation has 2.7 more pilots for every transmitter. 
\begin{figure}[h!]
\begin{center}
\begin{psfrags}
    \psfragscanon
\includegraphics[trim={0.1cm  0.3cm  0.1cm 1.3cm},clip,width=9cm]{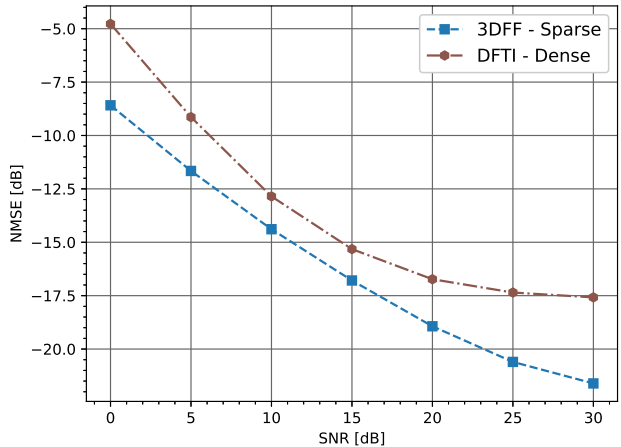}
\caption{\textbf{RS resource reduction}: 3DFF perform better than DFTI, despite the fact that 3DFF used sparse RS allocation (14\% overhead) while DFTI processed dense RS allocation (34\% overhead). \vspace{-0.3cm}} \label{8x8_3pVs8p}
\psfragscanoff
\end{psfrags}
\end{center}
\end{figure}

In Fig. \ref{8x8_3pVs8p} we see that the 3DFF estimator for sparse RS allocation performs better than DFTI for dense RS allocation, even though sparse allocation has 62.5\% less reference signals.

\begin{figure}[h!]
\begin{center}
\begin{psfrags}
    \psfragscanon
\includegraphics[trim={0.1cm  0.3cm  0.1cm 1.2cm},clip,width=9cm]{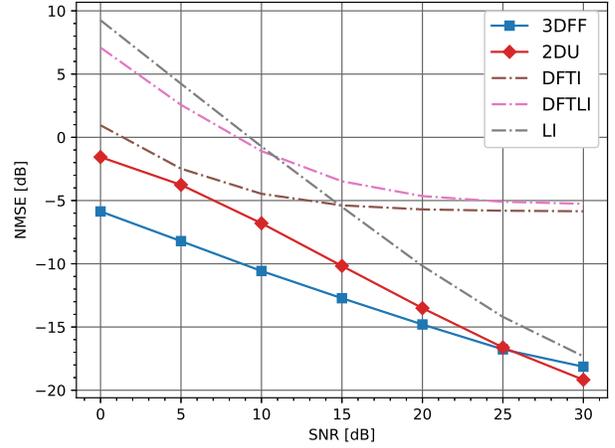}
\caption{\textbf{16x16 MIMO}: CE performance of 16x16 MIMO with sparse RS allocation. The 3DFF estimator achieves much better results, especially for medium SNR levels.\vspace{-0.5cm}} \label{16x16_3p}
\psfragscanoff
\end{psfrags}
\end{center}
\end{figure}

This is even more important for higher MIMO setups, where dense allocation is not feasible due to its high resource overhead. Therefore we have used a sparse RS allocation for training 16x16 MIMO. Fig. \ref{16x16_3p} shows that both CNN estimators perform better than the model-driven estimators, and the 3DFF estimator achieves significantly better results. Moreover, it is evident that as the MIMO scale increases, the performance gap between the CNN and the model-driven estimators increases in favor of the CNN estimators. 




\subsection{Combining Channel Estimation with MIMO Detection}
In this section we show the effect of the proposed CNN estimators on MIMO detection BER performance. We compare the LI estimator and our proposed CNN estimators by transmitting QPSK modulated data in a 8x8 MIMO, dense RS setting.
\begin{figure}[h]
\begin{center}
\begin{psfrags}
\includegraphics[trim={0.1cm  0.1cm  0.1cm 0.6cm},clip,width=9cm]{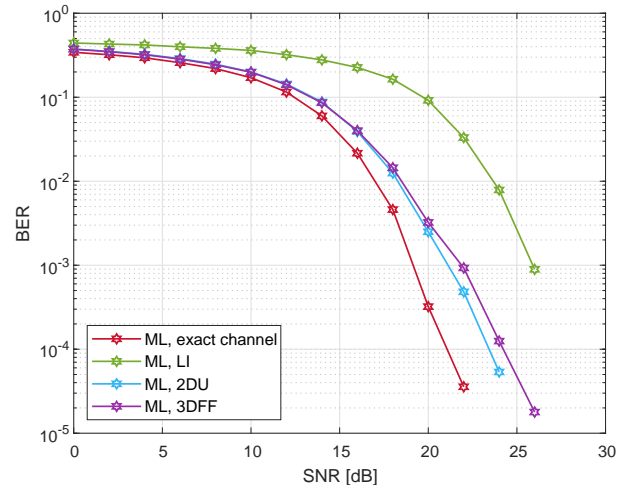}
\caption{\textbf{Maximum Likelihood (ML) MIMO detection:} The BER performance of our CNN estimators is considerably better than the LI estimator.\vspace{-0.2cm}} \label{ML_detection}

\end{psfrags}
\end{center}
\end{figure}
Transmissions are done with low correlation and no Doppler shift, so the LI estimator should perform relatively well, and the 3DFF estimator should have no advantage over the 2DU estimator. In Fig. \ref{ML_detection} we test the channel estimators with the optimal Maximum Likelihood (ML) MIMO detection algorithm. The BER achieved with both CNN estimators is much closer to the BER achieved when the exact channel is used for MIMO detection than that of the LI estimator.

Although the ML is an optimal detector, it is impractical for higher MIMO scales due to its exponential computational complexity. We therefore tested our estimators using 3 additional MIMO detection methods: Zero Forcing (ZF), V-BLAST and a NN MIMO detector which applies an iterative unfolding algorithm as described in \cite{NN_detector}. 
\begin{figure}[h]
\begin{center}
\begin{psfrags}
\vspace{-0.2cm}
\includegraphics[trim={0.1cm  0.3cm  0.1cm 0.8cm},clip,width=9cm]{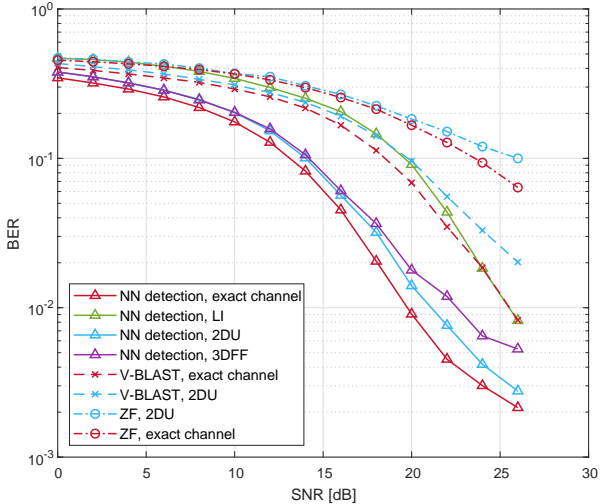}\vspace{-0.1cm}
\caption{ \textbf{Practical MIMO detection comparison:} Our 2DU estimator performance is almost as good as using the exact channel in all 3 practical MIMO detection methods.} \label{NN_detection}
\end{psfrags}
\end{center}
\end{figure}
\vspace{-0.2cm}
In Fig. \ref{NN_detection} we show that the NN detector trained with the output of our CNN estimators outperforms the other classic methods. Furthermore, the NN detector was shown in \cite{NN_detector} to achieve good performance with analog impairments and coded data. The fact that both of these algorithms are NN-based, and each algorithm by itself presents attractive advantages, motivates us to further research a joint NN architecture that includes both CE and MIMO detection.

\section{Conclusions and Future Work \label{s_conclusion_future}}
In this study we have presented a novel application of 2D U-net and 3D CNN architectures to the problem of MIMO-OFDM channel estimation. We have shown empirically that our 2DU and 3DFF CNN estimators outperform the model-driven estimators LI, DFTI, and DFTLI. Furthermore, the estimators' generalization capacity is evident by the usage of a highly diverse test set with various channel parameters. We also showed that using the 3DFF estimator can decrease the pilot resource allocation overhead by 62.5\%. Finally, We showed that the estimations provided by the CNN estimators result in good MIMO detection BER performance. 

Future work will involve expanding the proposed CNN estimators using RNN (Recurrent Neural Network), and specifically LSTM (Long Short-Term Memory) in order to exploit longer time evolution patterns of the wireless channels. 
This extension should further reduce the overhead of RS allocations without decreasing the CE performance. 
In addition, we aim to integrate the CE and the MIMO detection neural networks, expecting to yield even better performance for future wireless communication networks.

\bibliographystyle{./bibliographh/IEEEtran}

\bibliography{./bibliographh/IEEEabrv,./bibtex.bib}
\end{document}